\RequirePackage{fix-cm}
\documentclass[smallextended]{svjour3}       
\smartqed  
\usepackage{graphicx}
\usepackage{amssymb}
\usepackage[cmex10]{amsmath}
\usepackage{algorithm}
\usepackage{algorithmic}
\usepackage{url}

\usepackage{amsfonts}
\usepackage[small]{caption}
\usepackage{subfig}
\usepackage{multirow}

\usepackage{url}
\usepackage{multirow}
\usepackage{array}
\usepackage{rotating}
\usepackage{psfrag}
\usepackage{makecell}
\newtheorem{mydef}{Definition}
\usepackage{subfig}
\usepackage[export]{adjustbox}
\usepackage{url,color,xcolor,mdframed}

\begin{document}

\title{Role of Interdisciplinarity in Computer Sciences: Quantification, Impact and Life Trajectory
}

\titlerunning{Role of Interdisciplinarity in Computer Sciences}        

\author{Tanmoy Chakraborty}

\authorrunning{T. Chakraborty} 

\institute{Indraprastha Institute of Information Technology, Delhi (IIIT-D), India
}

\date{Received: date / Accepted: date}

\maketitle

\begin{abstract}
The tremendous advances in computer science in the last few decades have provided the platform to address and solve complex problems 
using interdisciplinary research. In this paper, we investigate how the extent of
interdisciplinarity in computer science domain (which is further divided into $24$ research fields) has changed over the 
last $50$ years. To this end, we collect
a massive bibliographic dataset with rich metadata information. We start with {\em quantifying}
interdisciplinarity of a field in terms of the diversity of topics and citations. We then analyze the effect of interdisciplinary research on the
scientific impact of individual fields and observe that highly disciplinary and highly interdisciplinary papers in general have a low
scientific impact; remarkably those that are able to strike a balance between the two extremes eventually land up having the highest impact. Further, we study the {\em reciprocity} among fields through citation interactions and notice that links from one field to related and {\em citation-intensive}  fields 
(fields producing large number of citations) are reciprocated heavily. A systematic analysis
of the citation interactions reveals the {\em life trajectory} of a research field, which generally undergoes three phases -- a {\em growing phase}, a {\em matured phase} and an {\em interdisciplinary phase}.  The
combination of metrics and empirical observations presented here provides general benchmarks for future studies of interdisciplinary
research activities in other domains of science.

\end{abstract}

\section{Introduction}

\emph{``Interdisciplinarity is viewed as a highly valued tool in order to restore the unity of sciences or to solve societal-pressing
	problems.''}

\hfill-- J. C. Schmidt \cite{Schmidt2008}\\

	With the advancement of scientific knowledge in a wide range
	of disciplines, researchers have started to increase their breadth of awareness  in order to answer difficult questions. 
	Therefore, many believe that today's research 
	takes place at the interstices between disciplines, and 
	does not follow any disciplinary boundaries \cite{Morillo,6734940}.  
	
	\if{0}
	Porter and Chubin~\cite{PorterC85} were the first who proposed that
	``Citations Outside Category'' can be a quite informative bibliometric measure. After that, many published attempts tried explain 
	interdisciplinarity \cite{klein,Metzger}. Rinia et al.~\cite{RiniaLBVR02} conducted an exploratory study of knowledge exchange between disciplines and subfields of
	science, based on bibliometric methods.  Further, they analyzed citation delay in interdisciplinary knowledge
	exchange within and across disciplines~\cite{Rinia}.  Urata~\cite{Urata90} made an attempt to identify the relationships among disciplines by examining the flow of citation. Steele and Stier~\cite{Steele:2000} used citation analysis and ordinary least squares regression to investigate the relationship between an article's
	citation rate and its degree of interdisciplinarity. 
	\fi

	It is evident that researchers conducting interdisciplinary research have
	access to more diverse
	opinions and patterns of thinking, which allow them to synthesize ideas from multiple disciplines and ask better research questions.
	However,
	few studies suggest that interdisciplinary research performs poorly based on many of the common metrics used to judge the quality of scientific research, including fewer citations per paper  and lower journal impact factors~\cite{Carayol01042005}. 
	There are however contradictory studies
	suggesting that there is no difference in average success of purely disciplinary and purely interdisciplinary papers~\cite{Jian}.  However, it has been repeatedly mentioned that there is a lack of proper quantitative indicator to measure interdisciplinarity ~\cite{Morillo,hiroki,caro}. Therefore, more research is needed
	to characterize interdisciplinary research in relation to disciplinary research, particularly in regard to the
	success of the research as a whole.

	Understanding the factors that facilitate interdisciplinarity are useful to get best integration to foster the evolution of new fields of scientific research. Although different interdisciplinarity
	measures have been proposed \cite{RiniaLBVR02,Urata90,Steele:2000}, none of these measures have been accepted for policy making purposes such as recruitment, fund disbursement etc.

	In this paper, we attempt to quantify the extent of interdisciplinarity of research areas in computer science domain and how this quantify changes over time  over the last 50 years. In particular, our work builds on the four fundamental questions pertaining to interdisciplinarity:
	\begin{itemize}
		\item {\bf Q1:} {\bf [Quantification]}  How to design suitable metrics to quantify the extent of interdisciplinary research in computer science?
		\item {\bf Q2:} {\bf [Impact]} What is the effect of interdisciplinary research on the overall scientific impact of any field in computer science?
		\item {\bf Q3:} {\bf [Reciprocity]} How does the interdisciplinary nature of a field influence for the reciprocation of citations? 
		\item {\bf Q4:} {\bf [Life Cycle]} Is it possible to describe the ``trajectory of life'' of a research field in terms of disciplinary and interdisciplinary activities inside that field? 
	\end{itemize}
	
	To this end, we collected a massive publication dataset related to computer science domain containing more than 2 million scientific articles enriched with metadata
	information. The entire computer science dataset is categorized into 24 research fields (Section \ref{dataset}). We start by showing evidences pertaining to the
rising pattern of interdisciplinary research over the years (Section \ref{evidence}). We then
propose two metrics to quantify the extent of interdisciplinarity of a research field (Section \ref{quantification}). We further show that research fields with a balanced degree of interdisciplinarity are more impactful compared to the fields with too
disciplinary or too interdisciplinary activities (Section \ref{impact}). Following this, we
study the ``reciprocity" of a research field in computer science domain and show that a field referring to the highly citation-intensive field is highly likely
to receive citations in reverse (Section \ref{Reciprocity}). Finally, we unfold the life trajectory of a research field which generally goes through three phases - a growing
phase, a matured phase and an interdisciplinary phase (Section \ref{life}).
	
 \if{0}   
    
	{\bf Evidences of Interdisciplinarity:} To begin with, we present a set of evidences to motivate the readers that over the years, interdisciplinary research in computer science is on the rise and presently it has become extremely hard to draw crisp boundaries among different research fields. We observe that the number of papers in relatively interdisciplinary fields as well as the diversity of contents in individual papers is accelerating at a faster rate than those in more disciplinary fields. At the same time, papers tend to cite other papers from diverse fields. We also observe strong indication about the increasing trend  of cross-field and inter-institution collaborations  (Section \ref{evidence}). 
	
	{\bf Quantification of Interdisciplinarity:} An important assumption of this study is that the references account for the relevance of the cited paper to the citing paper. Therefore, cross-field
	references in scientific publication may give a partial indication
	of knowledge transfer between fields within a domain. To answer our first question how to quantify the interdisciplinary of a research field, we therefore suggest a metric, called {\em Reference Diversity Index} (RDI). Furthermore, we capture the diversity of the content in terms of the associated keywords of papers and suggest another metric, called {\em Keyword Diversity Index} (KDI). Measuring and ranking research fields based on these metrics in different time windows reveal two interesting outcomes -- (i) all the fields show a consistent trend towards increasing  interdisciplinarity; (ii) the ranking of fields in terms of interdisciplinarity seems to change drastically over time -- fields like World Wide Web, Data Mining, Natural Language Processing, Computational
	Biology, Computer Vision gradually move towards the top position and Algorithms
	and Theory, Programming Languages, Operating Systems shift towards the bottom of the rank list. An immediate question that stems up is whether a field needs to necessarily promote interdisciplinary research to enhance its scientific impact? Our analysis reveals that for the entire computer science domain, in general, highly disciplinary and highly interdisciplinary research imply low scientific impact as compared to those which have a more balanced mix. However, for each individual field
	it is difficult to find any
	correlation between the extent of interdisciplinarity of papers and their
	scientific impact. But, there are few fields for which the level of interdisciplinarity and citation rates highly correlate with each other. 
	For the remaining fields, citations decline as interdisciplinarity grows (Section \ref{quantification}).

	{\bf Impact of Interdisciplinarity:} To answer the second question, we use three citation-based indicators -- citations per paper, Journal Impact Factor and most-cited papers per field. We observe that more disciplinarity and interdisciplinarity lead to low scientific impact as compared to the case when there is an equal mix. We further analyze the submission and acceptance statistics of top conferences in different fields. We observe that the interdisciplinary conferences such as WWW, ICDM, CVPR become extremely competitive in terms of high submission rate and less acceptance rate compared to disciplinary conferences such as STOC, FOCS etc (Section \ref{impact}).
	
	{\bf Reciprocity among research fields:} To answer our third question, one could relate the difference between fields in terms of their extent of interdisciplinarity and scientific impact with the intrinsic characteristics of fields being cited. We can explain it by the fact that few areas are more ``citation-intensive'' (fields tend to produce large number of citations) than others. Therefore, papers having a large number of interdisciplinary links with those citation-intensive fields might expect a lot of citations from them. Therefore, a field in general, might intend to get initial attention from the citation-intensive fields by citing them first with the anticipation of obtaining more citations in return from them. This phenomenon is known as ``reciprocity''  in network science. We observe that although the overall reciprocity of computer science domain is low, this tendency is significantly high among the related research fields, i.e., papers referring to the highly citation-intensive fields are highly likely to get citations in reverse from those fields (Section \ref{Reciprocity}).

	{\bf Life Trajectory of a research field:} To answer the final question, we  explore the ``trajectory of life''
	of a research field using simple
	bibliographic indicators. A case study on Data Mining (which has long temporal
	bibliographic evidences in our dataset) reveals that a field in general
	goes through three phases -- a {\em growing phase}, a {\em matured phase} and an {\em interdisciplinary phase}. In the growing phase, the field accumulates ideas from other fields. In the matured phase, the field produces many in-house citations. It also starts receiving citations from other fields. In the interdisciplinary phase, the field receives myriad of citations from other fields. The mutual interactions among many such fields may in turn create a completely new field (Section \ref{life}). 

}

\begin{figure}[!htb]
  \begin{mdframed}[backgroundcolor=gray!10]
  This measurement study unfolds the interdisciplinary activities
  in computer science over the last 50 years, pointing to microscopic effects causing the acceleration of the growth of this domain. \\
  {\color{blue}{\bf Key Insights:}}\\
  $\bullet$ Interdisciplinary research in computer science is increasing over the years.\\
  $\bullet$ Extensive interdisciplinarity might dilute the originality of a field; therefore a field should envisage to strike a balance between disciplinary and interdisciplinary research in order to achieve higher scientific impact.\\
  $\bullet$ Citations among a related set of fields are often reciprocated; in general, however, citations are rarely reciprocated among a random set of unrelated fields.\\
  $\bullet$ The ``life trajectory'' of a research field usually passes through three phases --  a growing phase, a matured phase and an interdisciplinary phase. 
   
  \end{mdframed}
\end{figure}

\fi

\if{0}
\begin{figure}[!htb]
  \begin{mdframed}[backgroundcolor=gray!10]
  This measurement study unfolds the interdisciplinary activities
  in computer science over the last 50 years, pointing to microscopic effects causing the acceleration of the growth of this domain. \\
  {\color{blue}{\bf Key Insights:}}\\
  $\bullet$ Interdisciplinary research in computer science is increasing over the years.\\
  $\bullet$ Extensive interdisciplinarity might dilute the originality of a field; therefore a field should envisage to strike a balance between disciplinary and interdisciplinary research in order to achieve higher scientific impact.\\
  $\bullet$ Citations among a related set of fields are often reciprocated; in general, however, citations are rarely reciprocated among a random set of unrelated fields.\\
  $\bullet$ The ``life trajectory'' of a research field usually passes through three phases --  a growing phase, a matured phase and an interdisciplinary phase. 
   
  \end{mdframed}
\end{figure}

\fi

\section{Related work}
In last few decades, researchers from different disciplines attempted to analyze the structural and dynamical properties of the citation and collaboration networks \cite{newman2,newman,price65}. At the same time, there has been attempts to quantify
interdisciplinarity of
scientific journals and researchers~\cite{Leydesdorff,alan,ismael}.  Porter and Chubin \cite{PorterC85} were the first who proposed that
``Citations Outside Category'' can be a quite informative bibliometric measure. Following this, many attempts further divided interdisciplinarity into components such as pluridisciplinarity, crossdisciplinarity, and even
metadisciplinarity~\cite{klein,Metzger}.
Rinia et al. \cite{RiniaLBVR02} studied how knowledge exchange happens between difference fields of
science based on bibliometric methods.  Further, they  analyzed citation delay in interdisciplinary knowledge
exchange within and across disciplines \cite{Rinia}.  Urata \cite{Urata90} made an attempt to identify the relationships among disciplines by
examining the flow of citation and the migration of scholars. Steele and Stier 
\cite{Steele:2000} used citation analysis and ordinary least squares regression to measure the correlation between the citation rate and the extent of interdisciplinarity. 
Morillo et al. \cite{ASI:ASI10326} measured
interdisciplinarity through a series of indicators based on Institute for Scientific Information (ISI) multi-assignment of journals in
subject categories. Then they establish a tentative typology of disciplines and research areas according to their degree of
interdisciplinarity. Levitt and Thelwall \cite{LevittT08} used count and diversity of citations to measure and compare the impact of journals classified in multiple subjects and journals classified in a single subject.

Pan et al.~\cite{sitabra} studied different fields of Physics (identified by different PACS codes) and showed a clear trend 
towards increasing interactions between different fields. 
They
concluded that due to the lack of citation information, they were unable to capture the micro dynamics controlling the
inherent interaction patterns among the fields.

In this paper, we extend our earlier study \cite{0002KRKGM13} where we built an automated system to classify core and interdisciplinary fields in computer science domain. We further studied the evolution dynamics at a microscopic level to show how interdisciplinarity emerges through
cross-fertilization of ideas between the fields that otherwise have little overlap as they are mostly studied independently. 
The major differences of the current paper from \cite{0002KRKGM13} are as follows: (i) a new bibliographic dataset is introduced; (ii) two new metrics measuring interdisciplinarity of a research field are proposed; (iii) we show that a coherent analysis of both the citations and references of a field can unveil the temporal evolution and the life cycle of a field. 



\section{Massive publication dataset}\label{dataset}
 We crawled Microsoft Academic Search (MAS), one of the publicly available repositories and collected all the papers related only to the computer science domain\footnote{The crawling process was completed in August, 2013.}. MAS is a semantic search engine, not a keyword-based one. Traditional search engines rely mostly on keyword matching. MAS is different from other academic search engines because it employs natural language processing; for example, the query ``machine learning''. It is possible that many such publications may not even include the words ``machine'' and ``learning'' in their titles or even text body. MAS dataset has been used in many previous studies \cite{Sinha:2015,Chakraborty:jcdl,Singh:2015:CCR,Wade:201,Chakraborty:2015:,7113314,PradhanPMNC17}. A detailed description of the dataset can be available in \cite{Chakraborty:jcdl}.

 The crawled dataset contains more than 2 million distinct papers. Moreover, MAS divided Computer Science domain into 24 fields and assigned each paper to one or more such fields (see Table \ref{table:field} for 24 fields and the percentage of papers per field). There are $8.68\%$ papers which are tagged with multiple fields (i.e., they belong to multiple fields).
 Each paper comes along with various bibliographic information -- title, a unique index, its author(s), year of publication, publication venue, its related field(s), abstract and keywords. Note that each paper was already annotated by MAS with a set of keywords to characterize the paper. These keywords are not given by the authors of the papers, but automatically extracted by MAS. We used these keywords further for our experiments. 

Each entry (corresponding to a paper) is shown in Figure \ref{entry}. Apart from the metadata information of all the papers, another advantage of using this dataset is that the ambiguity of named-entities (authors and publication venues) has been completely resolved by MAS, and a unique identity is associated with each author, paper and publication venue \cite{Chakraborty:jcdl}.  However, since we collected only Computer Science related papers, there is no evidence that a paper is assigned also to other non-Computer Science fields.

\begin{figure}[!htb]
  \begin{mdframed}[backgroundcolor=blue!10]
\#*GlitchMap: An FPGA Technology Mapper for Low Power Considering Glitches.\\
\#@Lei Cheng,Deming Chen,Martin D. F. Wong\\
\#t2007\\
\#cDAC\\
\#fComputer Architecture\\
\#kField programmable gate arrays, Minimization methods, Delay, Table lookup, Energy consumption, Power engineering computing, Algorithm design and analysis, Boolean functions, Permission, Logic\\
\#index134672\\
\#\%233644\\
\#\%759\\
\#\%283365\\
\#\%215199\\
\#\%282586\\
\#\%214457\\
\#\%132100\\
\#\%281965\\
\#\%281805\\
\#!In 90-nm technology, dynamic power is still the largest power source in FPGAs [1], and signal glitches contribute a large portion of the dynamic power consumption. Previous power-aware technology mapping algorithms for FPGAs have not taken into account the glitch power reduction. In this paper, we present a dynamic power estimation model and a new technology mapping algorithm considering glitches. To the best of our knowledge, this is the first work that explicitly reduces glitch power during technology mapping for FPGAs. Experiments show that our algorithm, named GlitchMap, is able to reduce dynamic power by 18.7\% compared to a previous state-of-the-art power-aware algorithm, EMap [2].
  \end{mdframed}
  \caption{An entry of the MAS dataset. Different tags indicate different attributes of the paper -- \#* is the title, \#@ is the author list, \#t is the year of publication, \#f is the related field of the paper, \#k is the set of keywords, \#index is the unique index of the paper, \#\% is the index of the paper which the current paper refers to, and \#! is the abstract of the paper. }\label{entry}
\end{figure}

\begin{table}[h!]
\begin{center}
\caption{Percentage of papers in various fields (with abbreviation) of computer science domain.}\label{table:field}
\scalebox{0.8}{
\begin{tabular}{|c|c|c|c|}
\hline
 Field & \% of papers & Field  &\% of papers\\ \hline
Artificial Intelligence (AI) & 12.64 & Algorithm (Algo) & 9.89 \\ 
Networking (NETW) & 9.41 & Databases (DB) & 5.18 \\ 
Distributed Systems (DIST) & 4.66 & Comp. Architecture (ARC) & 6.31 \\ 
Software Engineering (SE) & 6.26 & Machine Learning (ML)  & 5.00 \\               
Scientific Computing (SC)) & 5.73 & Bioinformatics (BIO) & 2.02 \\ 
Human Comp. Interaction (HCI) & 2.88 & Multimedia (MUL) & 3.27 \\ 
Graphics (GRP) & 2.20 & Computer Vision (CV) & 2.59 \\ 
Data Mining (DM) & 2.47 & Programming Language (PL) & 2.64 \\ 
Security and Privacy (SEC)  &2.25 & Information Retrieval (IR) & 1.96 \\ 
Natural Language Processing (NLP) & 5.91 & World Wide Web (WWW) & 1.34 \\
Education (EDU) & 1.45 & Operating Systems (OS) & 0.90 \\ 
Real Time Systems (RT) & 1.98 & Simulation (SIM) &1.04 \\ \hline
\end{tabular}}
\end{center}
\end{table}

\begin{figure}[!h]
\centering
 \includegraphics[width=\columnwidth]{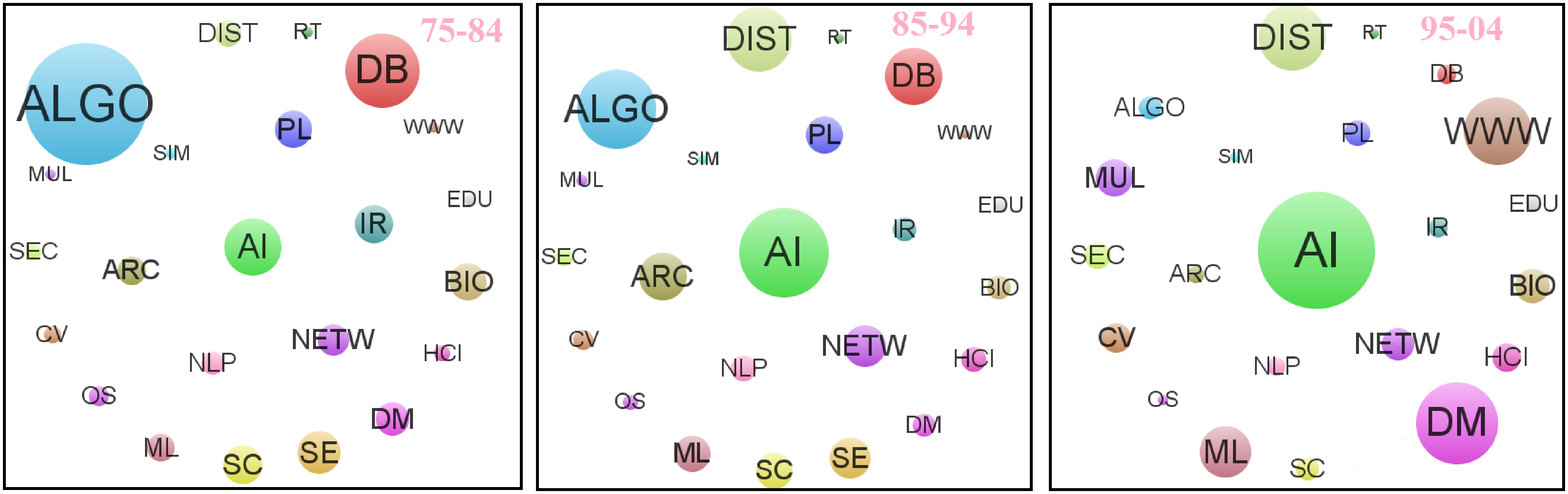}
 \caption{Number of papers in each field in three successive decades (1975-1984, 1985-1994 and 1995-2004). The
size of each circle is proportional to the
number of papers published in the corresponding field.
}\label{paper}

\end{figure}

\section{Evidences of interdisciplinarity}\label{evidence}
One of the simplest evidences of rising popularity of interdisciplinary research can be observed from the growth of the number of interdisciplinary papers over the years. In Figure~\ref{paper}, we observe that in the initial years (1975-1984), the fields like
Algorithms,
Databases seemed to have fully dominated the computer science research; however the trend has gradually shifted with the appearance of fields
such as Distributed Systems, Networking and Computer Architecture in the middle of 80's. In the recent decade, while the number of papers in
the fields like Algorithms, Databases, Operating Systems 
tend to diminish significantly, the
relatively new areas such as WWW, Data Mining,
Multimedia show a larger volume of publications. This result presents a prefatory evidence of the increasing research in the interdisciplinary fields vis-a-vis a decreasing trend of research in the disciplinary fields.

\begin{figure}[!ht]
\centering
 \includegraphics[width=\columnwidth]{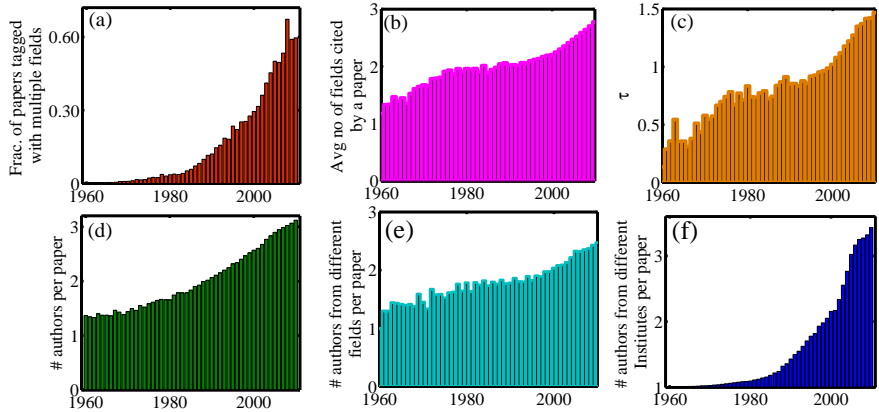}
 \caption{Preliminary evidences (upper panel: paper-centric, lower panel: author-centric) of increasing interdisciplinarity
in computer
science domain -- with the increase in time, (a) fraction of
papers tagged by multiple fields (i.e., papers belonging to more than one field among 24 fields in Computer Science), (b) average number of fields cited by a paper in its reference section, (c) $\tau$ (ratio
between cross-field and self-field references given by a paper), (d) average team-size (number of authors) per paper, (e) average number of
authors from different fields per paper, (f) average number of authors from different institutes per paper.}\label{evi}

\end{figure}

Further, we use the metadata information of our dataset and present a series of six evidences to support the rising popularity of interdisciplinary research in the computer
science domain. The evidences are drawn from a high level analysis of the publication dataset. \\

\noindent{\bf Paper-centric evidences.} Figure~\ref{evi}(a) shows the number of papers tagged by multiple fields over
the years. We hypothesize that the related field of a paper shows its disciplinarity, and being tagged with multiple fields can be an
evidence of the interdisciplinary nature of the paper. Around 8.68\% papers are tagged with multiple fields (belong to more than one field among 24 fields in computer science), and the
increasing trend of such papers over the years indicates the rise of interdisciplinary research. Then we consider the references
of a paper and measure the average number of fields cited by the paper. This measure is an easy way to check from where the ideas of the examined paper has taken since the references of a citing paper are assumed to be the sources of knowledge of that paper. One can therefore loosely correlate this raw measure with interdisciplinarity. For each paper, we consider the set of its references and their
associated fields, and measure the number of different fields cited by the paper. The more the citing paper refers to papers from multiple fields, the more the paper tends
to
become interdisciplinary. Figure~\ref{evi}(b) shows a steady increase in the breadth of citing in different fields over the years. Next, we
measure the ratio between cross-field and same-field references per paper, indicated by $\tau$. 
The increasing trend in Figure~\ref{evi}(c) indicates that most of the research fields have already been applying ideas from the other fields, and this hybridization in turn can trigger the emergence of a new research field in the immediate future.\\

\noindent{\bf Author-centric evidences.} Here we start with measuring the effect of team-size (number of authors) per paper, which is often (and sometimes mistakenly) associated with interdisciplinarity. Figure~\ref{evi}(d) shows that in computer science, the number of authors per paper
has escalated remarkably, with about 75\% average growth.   In one of our previous
studies~\cite{snam}, we showed that not only has the team size increased over the years, but also multi-continent collaboration has
increased largely in the last two decades. These are some of the elementary evidences of hybridization of ideas that take place while
collaborating with different
researchers. In one step further, we see for each paper the average number of authors having expertise in different fields. In
Figure
\ref{evi}(e), we again witness a modest increase in the number of fields on which the authors of a paper have expertise in. The rich metadata
information further helps us to examine the inter-institution collaboration among researchers. In Figure~\ref{evi}(f), we observe an increasing trend in the number of papers written by authors from different institutes.  
All these evidences indeed attest to notable changes in research practices over the last 50 years in computer science domain.  

\begin{figure}[!htb]
  \begin{mdframed}[backgroundcolor=gray!10]
  {\color{blue} {\bf Findings 2:}}\\
    $\bullet$ Number of papers in relatively interdisciplinary fields is accelerating at a faster rate than those in more disciplinary fields.\\
    $\bullet$ The evidence of interdisciplinarity becomes prominent with the increasing number of papers having diverse content (in terms of keywords).\\     
    $\bullet$ The increasing tendency of a paper referring to papers in diverse fields serves as an indicator for interdisciplinarity.\\
    $\bullet$ Collaborations of researchers across fields and institutes should strengthen interdisciplinary activities.
  \end{mdframed}
\end{figure}

 \section{Methodology}\label{method}
 
\subsection{Quantifying interdisciplinarity}\label{quantification}

All the evidences demonstrated so far coincide to a common conclusion that more and more the domain is resorting
 to interdisciplinary research
because the problems at hand in computer science are now beyond the boundaries of any one single field. Although interdisciplinarity has been quantified using external attributes (e.g., research formulation, team processes,
collaborations, research outputs, dispersion)~\cite{caro}, here we concentrate on two intrinsic properties of a scientific paper -- references and content (in terms of keywords) \cite{6734940}. In particular, we measure how references emitting from a paper point to other fields, and how diverse are the keywords associated with the paper to quantify interdisciplinarity.  Let us consider $F=\{F_1, F_2, \cdots\}$, a set of fields (in our dataset, there are $24$ fields) and $P_i=\{P_i^1,P_i^2,\cdots\}$, a set of papers related to field $F_i$. $\mathcal{R}_p$ is the number of references of paper $p$  and $\mathcal{R}_p(F_j)$ indicates the number of references of $p$ pointing to the papers in field $F_j$. 

In bibliographic research, the references of a paper are assumed to be the indicators of the related subject areas from where the citing paper has been motivated. Moreover, it is quite intuitive that the more diverse the references of a paper, the more the probability that the paper falls in the interdisciplinary regime. We measure ``diversity'' by Shannon's entropy, which provides an indicator of (un)evenness. 
Then, we measure interdisciplinarity of a field by {\em Reference Diversity Index}.

\begin{mydef}
{\bf Reference Diversity Index (RDI).}  The RDI of a field $F_i$ is defined by the average entropy of references of its related papers pointing to all the fields:
\begin{equation}
 RDI(F_i)=\frac{1}{|P_i|}\sum_{p\in P_i} \sum_{j} - \frac{\mathcal{R}_p(F_j)}{\mathcal{R}_p}  \log \frac{\mathcal{R}_p(F_j)}{\mathcal{R}_p}  
\end{equation}
\end{mydef}

The next metric is formulated from the evidence of the keywords associated with each paper. In our dataset, MAS assigns keywords to each paper from a global set of keywords in order to characterize it's content. It has been already shown that keywords of a paper can be used systematically to measure how different a paper is from the rest of the lot \cite{Chakraborty:jcdl}. Here, we use this observation to formulate another metric, called {\em Keyword Diversity Index} to measure the interdisciplinarity of a paper vis-a-vis a field. For a field $F_i$, we collect all the keywords from the papers of $F_i$ and make a set $\mathcal{K}_{F_i}$. There might exist keywords which are part of multiple fields, and these are the indicative evidences of interdisciplinarity. Along with the earlier notations, let us further define  $\mathcal{K}_p$ to be the set of keywords of paper $p$. Then  Keyword Diversity Index is defined as follows.

\begin{mydef}
 {\bf Keyword Diversity Index (KDI).} The KDI of a field $F_i$ is defined by the average diversity of keywords associated with papers in field $F_i$:
 \begin{equation}
  KDI(F_i)=\frac{1}{|P_i|} \sum_{p\in P_i} \sum_{j} - \frac{|\mathcal{K}_{F_j} \cap \mathcal{K}_p|}{|\mathcal{K}_p|}  \log  \frac{|\mathcal{K}_{F_j} \cap \mathcal{K}_p|}{|\mathcal{K}_p|} 
 \end{equation}

\end{mydef}

\subsection{Citation-based impact measurement}\label{sec:impact}
Scientific impact measures can be formulated using the citation statistics. Here we use three measures~\cite{schubert_1986,Moed} to calculate the impact of a research field as follows:
\begin{itemize}
\item{\bf Citations per paper (CP):} For each paper, we consider the total number of citations within first 5-years of its publication (in order to consider ``aging phenomenon''). We exclude those citations where first author is common in both citing and cited papers \cite{pon}.
 \item {\bf Journal Impact Factor (JIF):} The impact factor of a journal in which the paper is published. 
Note that for a paper, the impact factor of a journal is calculated at the time of publication of the paper.
\item{\bf Most cited papers:} Percentage of papers of a field among the top 5\% most-cited papers of all the fields.
\end{itemize}

\section{Experimental Results}

\subsection{Interdisciplinary of research fields}
\begin{figure}[!ht]
    \centering
    \subfloat[]{{\includegraphics[width=\columnwidth]{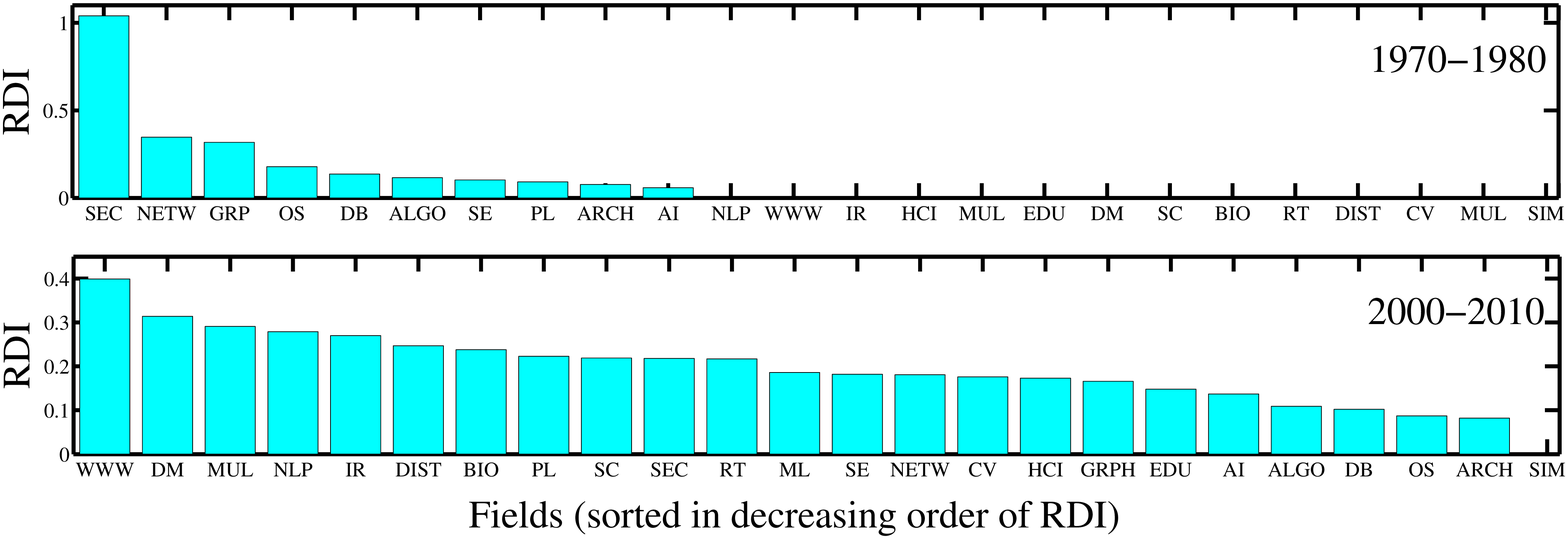}}}%
    \qquad
    \subfloat[]{{\includegraphics[width=\columnwidth]{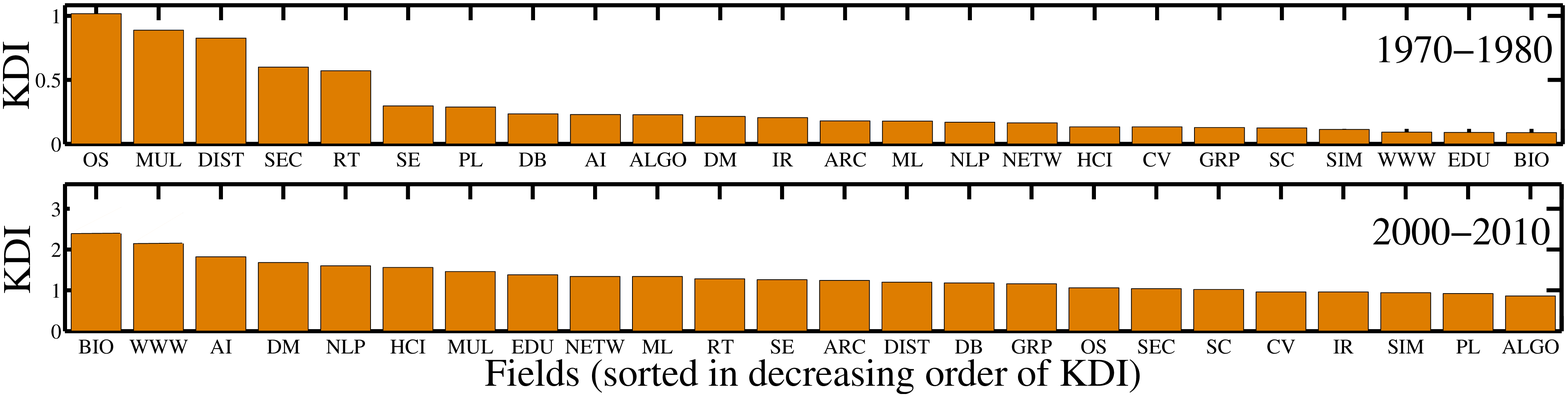} }}%
    \caption{Ranking of fields based on two interdisciplinarity measures in 1970-1980 and 2000-2010. In x-axis, the fields are sorted based on the measure in y-axis. We notice two important observations: (i) ranking of the fields has been changed drastically over the years, (ii) overall the interdisciplinarity values of most of the fields increase over time. }%
    \label{ranking}%
\end{figure}

We measure the interdisciplinarity of all the research fields in computer science domain using the suggested measures in different time periods. Figure~\ref{ranking} illustrates the ranking of the fields based on RDI and KDI in two decades. The more
the values of RDI and KDI of a field, the more it turns out to be an interdisciplinary field. We observe that the interdisciplinarity in computer science mostly started accelerating after 1980. Fields like Data Mining, World Wide Web, Human Computer Interaction, Bioinformatics hold a consistent position at the top based on the interdisciplinary values. On the other hand,  fields like Algorithms, Operating
Systems, Hardware and
Architecture, Databases, Programming Languages gradually shift towards the bottom of the 
rank. 
Surprisingly, we also observe that in general the extent of interdisciplinarity for most of the fields  tends to increase steadily (the interdisciplinarity scores, i.e., RDI and KDI, for most fields become almost double in 2000-2010 as compared to that in 1970-1980). This doubling is observed across almost all fields.
The results emphasize the fact that all the fields steadily become interdisciplinary, and therefore it is extremely difficult to differentiate one field from other in a domain.

\begin{figure}[!htb]
  \begin{mdframed}[backgroundcolor=gray!10]
  {\color{blue} {\bf Findings 3:}}\\
    $\bullet$ We measure interdisciplinarity by means of two intrinsic properties of a paper -- its references and the associated keywords.\\
    $\bullet$ The extent of interdisciplinarity of most of the fields tends to increase over time,  pointing to the fact that fields are getting more overlapped.
    \end{mdframed}
\end{figure}

\subsection{Correlation between interdisciplinarity and scientific impact}\label{impact}
A crucial question in scientific research is -- how does interdisciplinarity lead to gain scientific impact or vice versa.  
In this section, we use the metrics to measure the scientific impact presented in Section \ref{sec:impact} and develop a connection between interdisciplinarity and scientific impact.

\if{0}
\subsubsection{Citation-based impact measurement}
Scientific impact measures can be formulated using the citation statistics. Here we use three measures~\cite{schubert_1986,Moed} as follows:
\begin{itemize}
\item{\bf Citations per paper (CP):} For each paper, we consider the total number of citations within first 5-years of its publication (in order to consider ``aging phenomenon''). We exclude those citations where first author is common in both citing and cited papers \cite{pon}.
 \item {\bf Journal Impact Factor (JIF):} The impact factor of a journal in which the paper is published. 
Note that for a paper, the impact factor of a journal is calculated at the time of publication of the paper.
\item{\bf Most cited papers:} Percentage of papers of a field among the top 5\% most-cited papers of all the fields.
\end{itemize}
\fi

\begin{figure}[!ht]
    \centering
    \subfloat[]{{\includegraphics[width=\columnwidth]{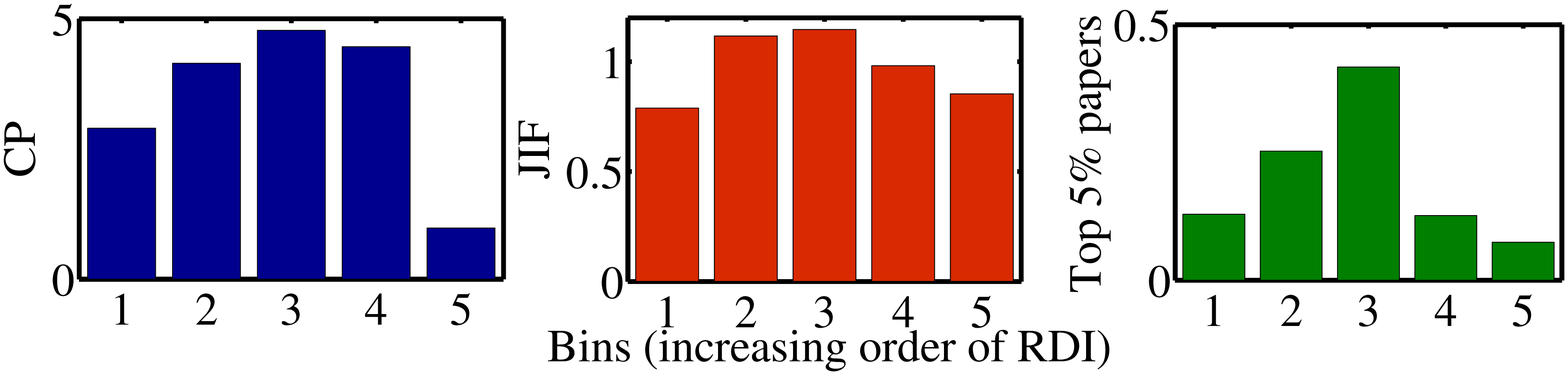}}}%
    \qquad
    \subfloat[]{{\includegraphics[width=\columnwidth]{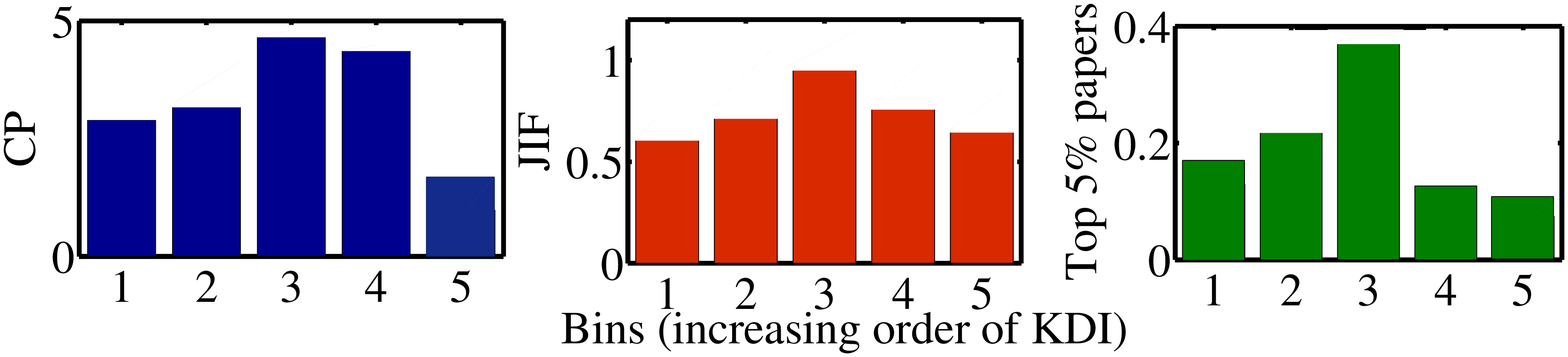} }}%
    \qquad
    \subfloat[]{{\includegraphics[width=\columnwidth]{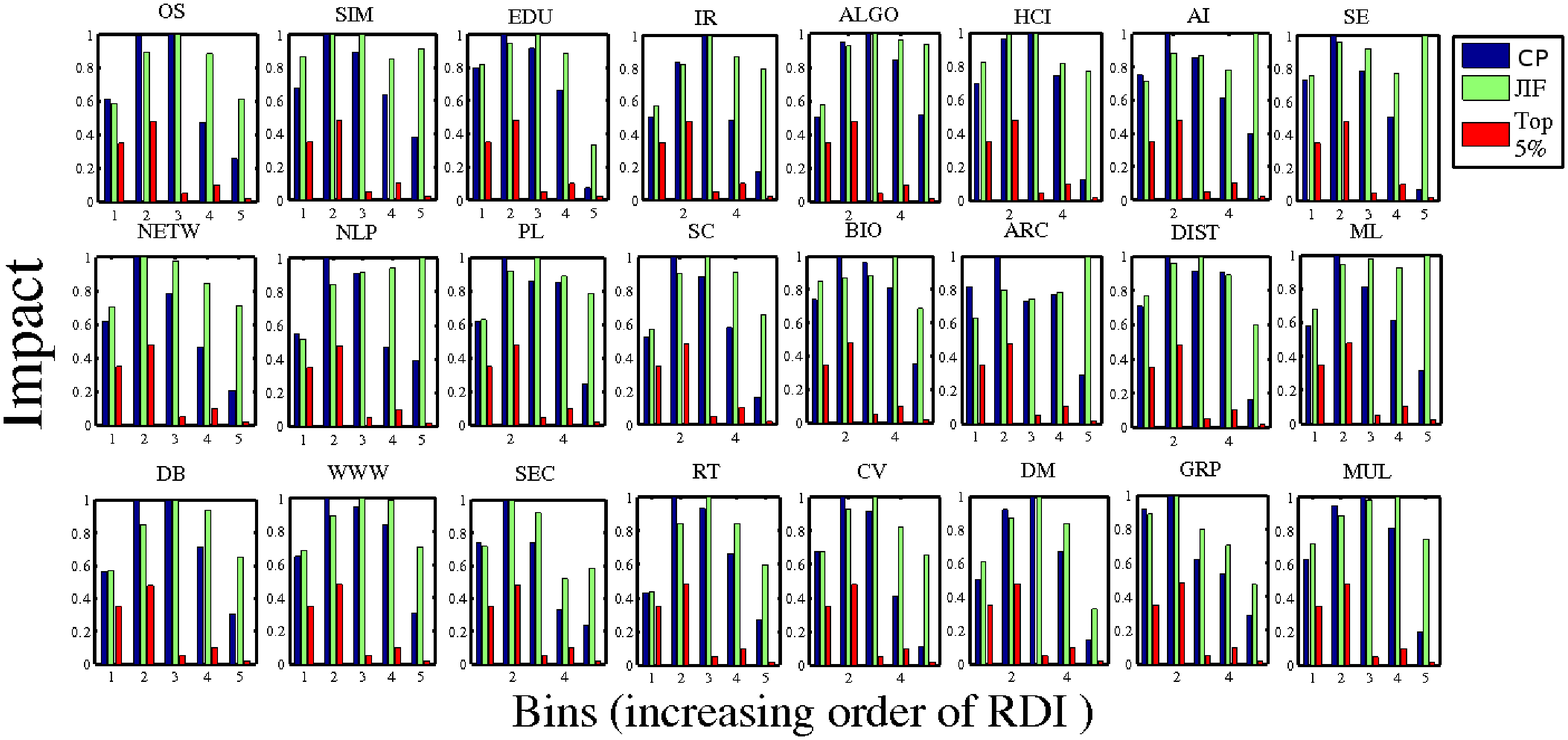} }}%
    \caption{Correlation between different interdisciplinary measures and the scientific impact for (a)-(b) overall computer science
domain and (c) all the fields. In x-axis, the range of each interdisciplinary measure is equally divided into five buckets (where bucket 1 ({\em resp.} 5) contains papers with low ({\em resp.} high) interdisciplinarity)  and in y-axis, we plot the average impact of the papers belonging to each bin.}%
    \label{impact_total}%
    
\end{figure}


Figures \ref{impact_total}(a)-(b) connect the interdisciplinarity and the scientific impact for the entire computer science
domain. For each interdisciplinary measure, we divide the entire range of the x-axis into five equal buckets where bucket 1 ({\em resp. } 5) contains papers with very low ({\em resp. } high) value of the metric (RDI/KDI) represented by the x-axis. For each bucket we measure the average
impact of the  papers falling in that bucket. 
It is evident from the figures that more disciplinarity and interdisciplinarity lead to low scientific impact as compared to the values at the middle range. Same conclusion can be drawn for all the fields as shown in
Figure \ref{impact_total}(c) -- purely disciplinary (lowest bucket) and purely interdisciplinary (highest bucket) papers in general exhibit: lower
citation rate, are published in lower impact factor journals
and are less likely to be amongst the 5\% most cited papers.  This result essentially indicates that the papers which exhibit high disciplinarity or high
interdisciplinarity are perhaps too narrow towards a specific field or too diverse to attract much citations compared to the papers which provide a mix of cited papers from different fields~\cite{LariviereG10}. However, the fields such as Software Engineering, Machine Learning,
NLP sometimes show contradictory behavior in the sense that the papers (related to these fields) which are highly interdisciplinary seem to
get published in high impact journals. However, their behavior for the other two impact measures follow the same general trend. The
reason could be that even if the journals where the highly interdisciplinary papers are published have high impact factors, the published
papers often fail to attract attention from diverse domains, resulting in deterioration in cumulative citation count.


 \subsection{Top-tier conference statistics}
\begin{figure}[!h]
\centering
\includegraphics[width=\columnwidth]{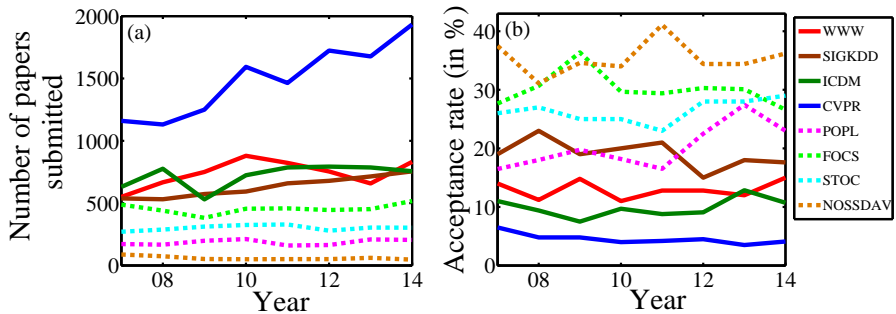}
\caption{Statistics of the top-tier conferences over the last eight
years -- (a) number of paper submitted and (b) acceptance rate of papers.}\label{conf}

\end{figure}

Another way of understanding the popularity of different research fields is to observe the (submission and acceptance) statistics of papers in top-tier conferences. To this end, we collected the statistics of four top-tier conferences identified by MAS each for disciplinary and interdisciplinary fields in computer science. The conferences related to disciplinary fields are -- STOC (Algorithm \& Theory), FOCS (Algorithm \& Theory),  POPL (Programming Languages) and NOSSDAV (Operating Systems). The conferences related to interdisciplinary fields are ICDM (Data Mining), SIGKDD (Data Mining),   WWW (World Wide Web) and  CVPR (Computer Vision). We mainly focus on the two dimensions of these conferences - (a) {\em productivity}, in terms of the number of papers submitted to the conferences, and (b) {\em competitiveness}, in terms of acceptance rate of papers in these conferences. In Figure~\ref{conf}, one can observe that for the interdisciplinary conferences, while the productivity tends to increase over time, the acceptance rate consistently remains quite lower as compared to the conferences related to the disciplinary fields. It clearly indicates that with the increasing trend of interdisciplinary research in the recent years, the  interdisciplinary venues gradually become extremely competitive  in terms of acceptance rate compared to the pure disciplinary venues.

\begin{figure}[!htb]
  \begin{mdframed}[backgroundcolor=gray!10]
  {\color{blue} {\bf Findings 3:}}\\
    $\bullet$ In general, papers with a balanced mix of disciplinarian and interdisciplinarity tend to attract more citations than papers with pure disciplinarity or pure interdisciplinarity.\\
    $\bullet$ However, few fields show contradicting trends -- NLP, ML exhibit high impact with high interdisciplinarity.\\ 
    $\bullet$ The challenge of publishing papers in interdisciplinary venues increases evenly with the productivity.
    \end{mdframed}
\end{figure}

\subsection{Reciprocity among different fields}\label{Reciprocity}
The difference in the rankings based on the interdisciplinarity among various fields could be related to the inherent
characteristics of the individual fields~\cite{WallaceLG09}. This might be explained in terms of the fact that some fields are
more {\em citation-intensive} (i.e., fields which tend to produce a lot of citations in general compared to the other fields). As a result, it might happen that papers having more interdisciplinary linkage with such fields may get more citations from the papers related to those fields. The basic idea is that if a paper $p$ from field $F_1$ cites more to the papers belonging to a citation-intensive field $F_2$, then there is a higher chance that in future papers in $F_2$ would cite $p$. This phenomenon is known as {\em reciprocity} in network
science~\cite{gl04}. In order to test this hypothesis, we require case by case analysis. Here we consider WWW and observe its related papers and their incoming and outgoing citation distributions. The top five fields (excluding WWW
itself) to which the papers of WWW emit maximum citations are: AI, ML, NLP, IR and DM. Now for each such highly-cited field $F$ (in this case, $F$ could be one of AI, ML, NLP, IR and DM), we divide the papers of WWW (published between 1990--1995) into two buckets: {\em Bucket-1:} these papers cite more than 50\% of times to the papers of field $F$, {\em Bucket-2:} these papers cite less than 50\% of times to the papers of field $F$. Then for each bucket, we measure ACP\footnote{ACP is measured by counting all the citations given by a set of papers and normalizing it by the number of papers in that set.}, the average number of citations given by a paper of $F$ to the papers in a given bucket. Table~\ref{acp} shows the size of the bucket and the corresponding ACP value. We observe that papers in {\em Bucket-1} exhibit significantly higher ACP value for each highly-cited field $F$ as compared to that in {\em Bucket-2}. The result is more prominent in the case of DM (papers in {\em Bucket-1}  have 126.36\% higher ACP value compared to {\em Bucket-2}), followed by NLP
(68.47\% higher) and IR (62.92\% higher). Interestingly, among these highly-cited fields, the most citation-intensive field (based on overall
ACP value) is DM, followed by AI, NLP, IR and ML -- this rank roughly correlates with the rank based on the ACP difference mentioned earlier.   
This analysis shows that papers referring to more citation-intensive fields are more likely to be cited back by those fields, resulting higher volume of citations compared to those papers referring to not so citation-intensive fields.

\begin{table}
\caption{ACP value of all the WWW papers (published between 1990-1995) due to the citations given by the top five highly-cited fields of
WWW.}\label{acp}
\centering
\scalebox{0.7}{
 \begin{tabular}{|c|c|c|c|c|}
\hline
Highly-cited & \multicolumn{2}{c|}{{\em Bucket-1}: More than 50\% times} & \multicolumn{2}{c|}{{\em Bucket-2}: Less than 50\% times}\\\cline{2-5}
field        &  Bucket size (in \%) & ACP value     & Bucket size (in \%) & ACP value \\\hline

  AI        &  23\%    & 4.72   & 77\% & 3.46 \\
  ML         &  36\%    & 3.42  &  44\%  & 2.23 \\
  NLP        & 21\%     & 4.65   & 79\%  & 2.76  \\
  IR         & 18\%     & 5.23  & 82\%  & 3.21  \\
  DM        & 27\%       & 4.98 &  73\% & 2.20  \\\hline
 \end{tabular}}

\end{table}

\begin{figure}[!ht]
\centering
\includegraphics[width=1.1\columnwidth]{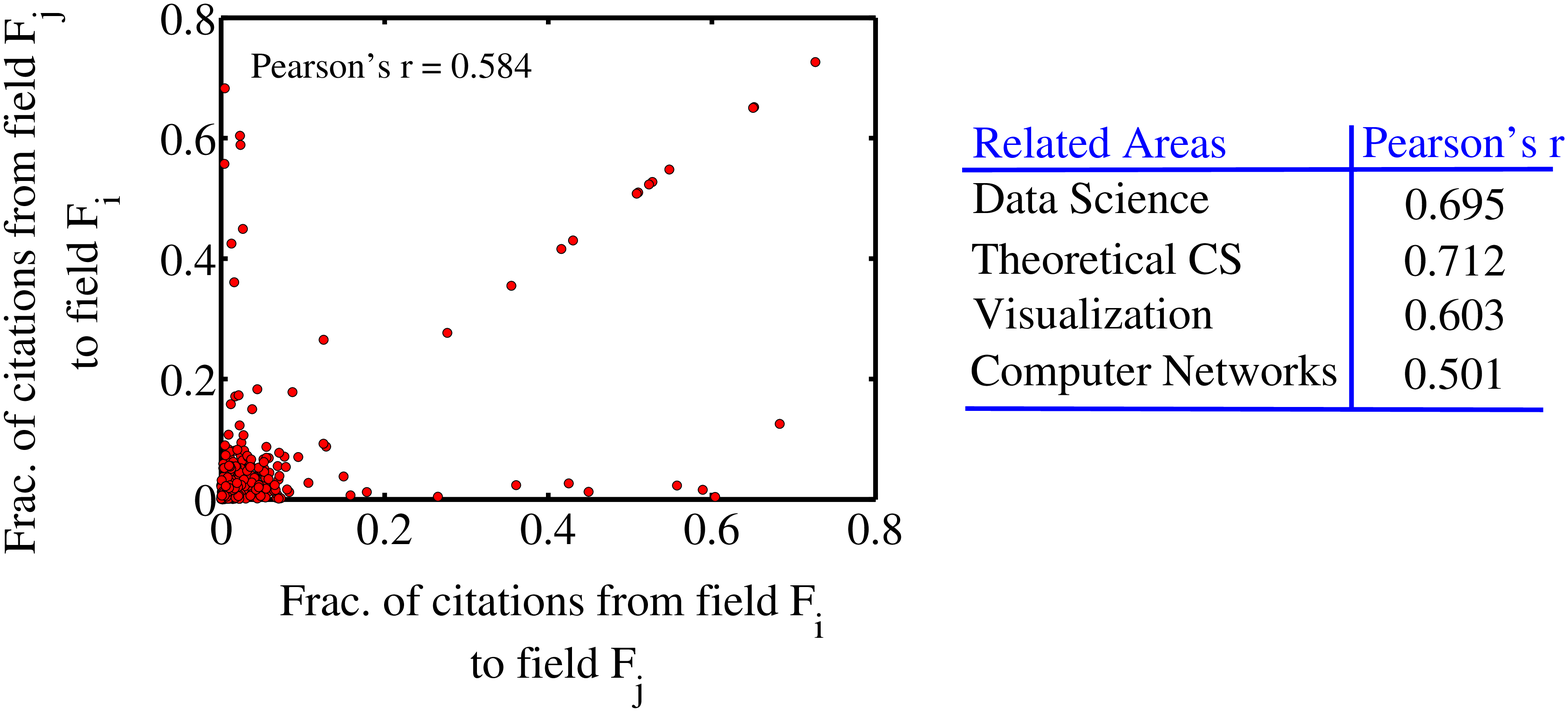}
\caption{Correlation between fraction of citations from field $F_i$ to field $F_j$ and that from field $F_j$ to field $F_i$.
Each point in the figure corresponds to the value for each pair of fields. Since we consider 24 fields, there are $24 \times 24$ points in
this scatter plot. We further measure the Pearson's correlation among the different related fields separately.}\label{cor}.

\end{figure}

At a broader level, we further measure how inbound and outbound citation counts between a pair of fields are correlated i.e.,  for each field the correlation between the percentage of its references made to
a field and the percentage of citations received from that field. More precisely, we intend to see the {\em reciprocity} of the entire computer science domain -- if papers in field $F_i$ produce
$x\%$ of citations for the papers in field $F_j$, will the papers in $F_i$ receive similar percentage of citations from the papers of $F_j$?
We measure the overall Pearson's correlation coefficient ($Pearson's\ r$) between these two entities for all pairs of fields and obtain a
reasonable value of $0.58$ as shown in Figure \ref{cor}. We anticipate that this correlation might be even better among the 
related fields. To this end, we further group the fields into four categories: (i) Data Science:  DB, DM, IR, NLP and ML, (ii)
Theoretical CS: ALGO, PL and SE, (iii) Visualization: GRP, CV, HCI and MUL, and (iv) Computer Networks: NETW, SEC, DIST and WWW. Note that there might be significant overlap across categories. For each
category, we measure $Pearson's\ r$ separately and observe that in these cases, the correlation is quite high -- the highest correlation is
obtained among the more disciplinary fields in computer science, i.e., Theoretical CS, which is followed by Data Science and visualization (see Figure \ref{cor}). This analysis
indeed unfolds a new perception on the citation dynamics among research areas --  if a field cites more to its related and citation-intensive field, in return it can expect high citations back from that field. This result might even have a stronger implication -- higher citations received by a field $F$ might not be only dependent on the intrinsic quality of the papers in $F$ but also on the interdisciplinary papers of certain other citation-intensive fields bringing in citations to $F$. Therefore, it would be worthwhile to look for a stronger measure that takes into account both the content of the paper and its interdisciplinary references to capture its quality and to predict its future importance.

\begin{figure}[!htb]
  \begin{mdframed}[backgroundcolor=gray!10]
  {\color{blue} {\bf Findings 4:}}\\
    $\bullet$ Papers referring to the highly citation-intensive fields are highly likely to get citations in reverse from those fields.\\ 
    $\bullet$ One should consider both the intrinsic quality of the paper as well as its interdisciplinary references to predict its future citations.
    \end{mdframed}
\end{figure}

\subsection{Life trajectory of a research field}\label{life}
\if{0}A presumption of this study is that references made to documented knowledge
account for the relevance of the previous work to the present research. Cross-field
citations in scientific and technical publications therefore, may give a partial indication
of knowledge transfer between fields of a domain.\fi Here, we present a concurrent analysis of the two citation components of a research paper: its incoming citations and its outgoing references. We hypothesize that both these components bear significant role in unfolding interesting aspects of a research field; one such aspect could be life trajectory of a field from its birth till death (if any). A full exploration of this aspect would require a very systematic and a case-by-case examination. To this end, we particularly focus on Data Mining (DM) which has sufficiently long history in our dataset. We explore the references emitting from the papers of DM and citations acquired by these papers over the years.

\begin{figure}[!t]
\centering
\includegraphics[width=\columnwidth]{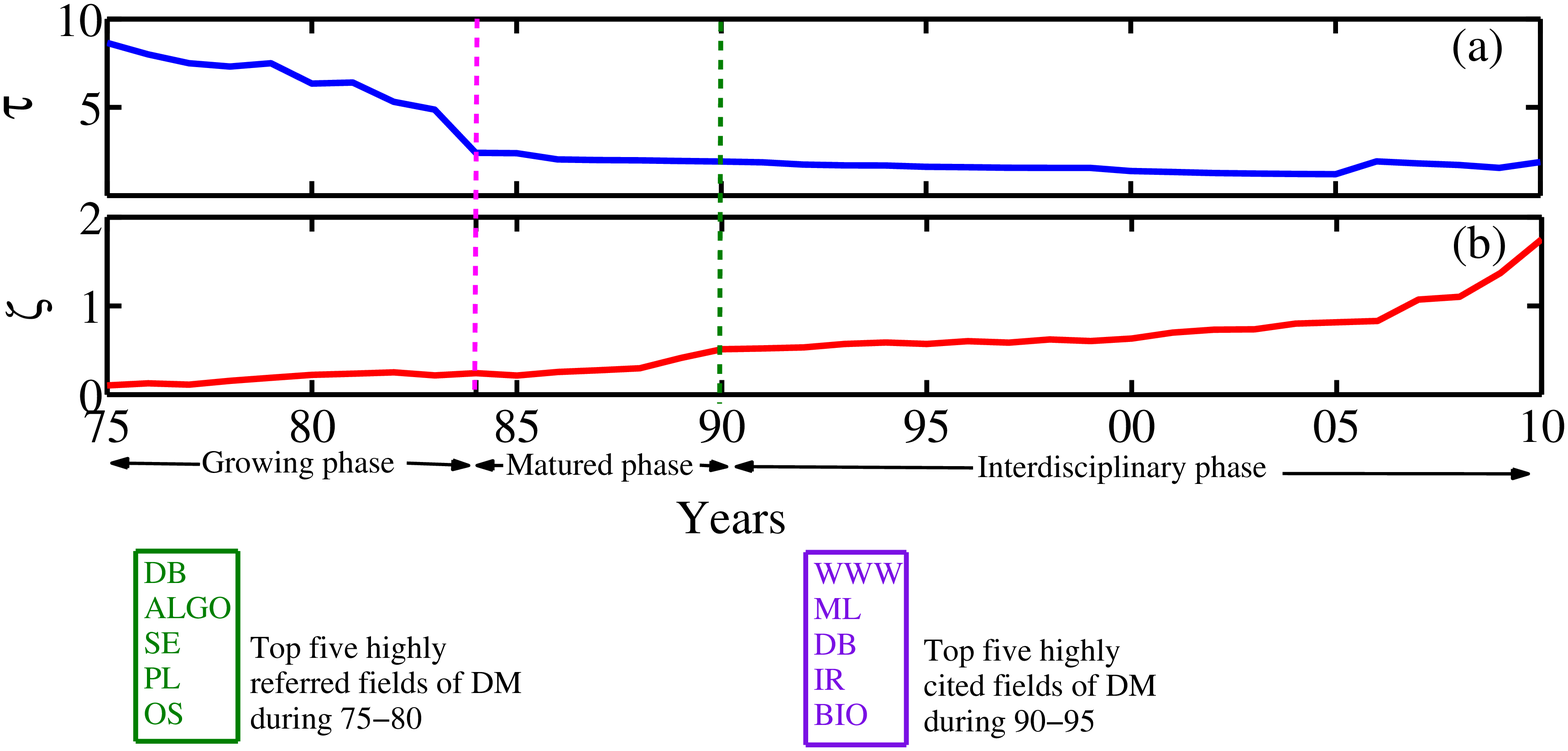}
\caption{ Life trajectory of a research field. Here we present the life trajectory of Data Mining (DM) passing through three successive
phases: Growing Phase, Matured Phase and Interdisciplinary Phase. We plot (a) $\tau$ (ratio between cross-field and same-field references
per paper in DM) and (b) $\zeta$ (ratio between the cross-field citations coming from other fields to DM and the same-field citations coming
from DM itself) over the years. We also show the top five highly referred fields of DM during 1975-1980 and top five highly-cited fields of
DM during 1990-1995.}\label{phase}

\end{figure}

We notice that in our dataset the number of papers in DM started increasing significantly from 1975. Therefore, we concentrate on all the
papers of DM from 1975 onward. We first plot $\tau$ (ratio between cross-field and same-field references per paper) for all the papers in
DM. As Figure~\ref{phase} shows, from 1975 to 1984 the value of $\tau$ is significantly high, indicating that during this time period DM
papers
mostly referred to the papers of other fields. This suggests that in this time window DM imported more knowledge from the other fields than
from itself.
 This phase can be considered as the ``growing phase''  of DM. The increasing number of
cross-field references indicates the acquisition of knowledge and ideas from different other fields which is necessary in the creation of a
new field. In Figure~\ref{phase}, we also show the top five fields which were mostly cited by DM. Databases and Algorithms
consistently remain two most-cited fields from where DM started gathering knowledge during its growing phase.
    
After 1984, the value of $\tau$ starts declining and remains almost stable till 2010. This indicates that after the birth of a field, it
tends to become matured enough to attract high number of self-field citations. However, $\tau$ refers to the distribution of references of
only DM papers. One might also be interested to look into the proportion of citations coming from other fields to DM papers. Therefore, we
plot $\zeta$ (ratio between the citations coming from other fields to DM and the citations coming from DM itself) for all the papers in DM. Self-field
citations  denote field's self-dependence -- fields with higher self-citation ratio tend to be more independent.
Interestingly, we observe that after 1990, $\zeta$ shows a rise, which indicates the acquisition of incoming citations of DM
papers from other fields. This period can be considered as the ``matured phase'' of DM. In Figure~\ref{phase}, we also present the top five
fields citing DM papers during 1990-1995. We notice that WWW and ML
are the two fields emitting maximum citations to DM during this time period. We also notice in our dataset that the growth of  WWW started from
1992 and it referred to DM and Networking papers most of the times. This perhaps indicates the birth of another field in the form of WWW
by the cross-fertilization of ideas coming from DM and other fields such as Networking and HCI. However, at 1990 DM seems to have become 
matured enough to  manifest new ideas that can in turn be transferred outside the field to trigger the birth of a new area of research in
computer science. One can consider this time as  the ``interdisciplinary phase'' of DM. Another evidence supporting the cross-fertilization of
DM and WWW is the increasing number of papers tagged by both DM and WWW present in our dataset. We observe 60\% increase of the papers
tagged by DM and WWW during 1990-1995 as compared to that in 1984-1989. During 1990-1995, given a paper tagged by multiple fields and one of such tagged fields is DM, the probability of the paper tagged also by WWW is 0.45, which is followed by ML (0.21) and DB (0.12).  
This analysis thus presents a landscape of the life trajectory experienced by a research field, which generally undergoes three phases -- a ``growing phase'', a ``matured phase'' and an ``interdisciplinary phase''. Although a universal acceptance of this trajectory model is far from reality and requires a
thorough understanding of the citation distribution of fields from other domains such as physics, life science, we believe that the
present observation can motivate researchers from other domains to explore the rise and fall of scientific paradigm in
a particular research area, and thus remains one of the immediate future directions to study.

\begin{figure}[!htb]
  \begin{mdframed}[backgroundcolor=gray!10]
  {\color{blue} {\bf Findings 5:}}\\
    $\bullet$ A research field generally undergoes three major phases -- a ``growing phase'', a ``matured phase'' and an ``interdisciplinary phase''.\\
    $\bullet$ In its ``growing phase'', a field accumulates ideas from other fields.\\ 
    $\bullet$ In its ``matured phase'', a field produces many in-house citations, i.e., citations pointing to the papers in that field itself. It also starts getting citations from other fields.\\ 
    $\bullet$ In its ``interdisciplinary phase'', a field receives myriad of citations from other fields. The mutual interactions among many such fields may in turn create a completely new field.\\ 
    \end{mdframed}
\end{figure}

\section{Discussion and future work}
In this article, we investigated how interdisciplinary research has evolved between 1960 and 2010 in the computer science domain in
general and its associated fields, including both old and established fields (e.g., Algorithms and Theory) and fields which are relatively new (e.g.,
World Wide Web). We suggested two metrics based on references and keywords to quantify the extent of interdisciplinarity. 
The observed results demonstrated that the practice of interdisciplinary research in computer science has undergone a modest increase.

Both qualitative and quantitative analysis on a huge bibliographic dataset unfolded several interesting inferences, many of which would have
not been exploited before due to the lack of appropriate knowledge base. Here, we summarize the major findings and discuss possible scopes of future research:

{\bf Evidences of interdisciplinarity:} We presented a set of evidences to motivate the readers that over the years, interdisciplinary research in computer science is on the rise and presently it has become extremely hard to draw crisp boundaries among different research fields. We observed that the number of papers in relatively interdisciplinary fields as well as the diversity of contents in individual papers is accelerating at a faster rate than those in more disciplinary fields. At the same time, papers tend to cite other papers from diverse fields. We also observed strong indication about the increasing trend  of cross-field and inter-institution collaborations. 

{\bf Quantification of interdisciplinarity:} An important assumption of this study is that the references account for the relevance of the cited paper to the citing paper. Therefore, cross-field
references in scientific publication may give a partial indication
of knowledge transfer between fields within a domain. To answer our first question how to quantify the interdisciplinary of a research field, we suggested a metric, called {\em Reference Diversity Index} (RDI). Furthermore, we captured the diversity of the content in terms of the associated keywords of papers and suggested another metric, called {\em Keyword Diversity Index} (KDI). Measuring and ranking research fields based on these metrics in different time windows revealed two interesting outcomes -- (i) all the fields show a consistent trend towards increasing  interdisciplinarity; (ii) the ranking of fields in terms of interdisciplinarity seems to change drastically over time -- fields like World Wide Web, Data Mining, Natural Language Processing, Computational
Biology, Computer Vision gradually move towards the top position and Algorithms
and Theory, Programming Languages, Operating Systems shift towards the bottom of the rank list. An immediate question that stems up is whether a field needs to necessarily promote interdisciplinary research to enhance its scientific impact? Our analysis revealed that for the entire computer science domain, in general, highly disciplinary and highly interdisciplinary research imply low scientific impact as compared to those which have a more balanced mix. However, for each individual field
it is difficult to find any
correlation between the extent of interdisciplinarity of papers and their
scientific impact. But, there are few fields for which the level of interdisciplinarity and citation rates highly correlate with each other. 
For the remaining fields, citations decline as interdisciplinarity grows.

{\bf Impact of Interdisciplinarity:} We used three citation-based indicators -- citations per paper, Journal Impact Factor and most-cited papers per field. We observed that more disciplinarity and interdisciplinarity lead to low scientific impact as compared to the case when there is an equal mix. We further analyzed the submission and acceptance statistics of top conferences in different fields. We observed that the interdisciplinary conferences such as WWW, ICDM, CVPR become extremely competitive in terms of high submission rate and less acceptance rate compared to disciplinary conferences such as STOC, FOCS etc.

{\bf Reciprocity among research fields:} One could further relate the difference between fields in terms of their extent of interdisciplinarity and scientific impact with the intrinsic characteristics of fields being cited. We can explain it by the fact that few areas are more ``citation-intensive'' (fields tend to produce large number of citations) than others. Papers having a large number of interdisciplinary links with those citation-intensive fields might expect a lot of citations from them. Therefore, a field in general, might intend to get initial attention from the citation-intensive fields by citing them first with the anticipation of obtaining more citations in return from them. This phenomenon is known as ``reciprocity''  in network science. We observed that although the overall reciprocity of computer science domain is low, this tendency is significantly high among the related research fields, i.e., papers referring to the highly citation-intensive fields are highly likely to get citations in reverse from those fields.

{\bf Life Trajectory of a research field:} We  explored the ``trajectory of life''
of a research field using simple
bibliographic indicators. A case study on Data Mining (which has long temporal
bibliographic evidences in our dataset) revealed that a field in general
goes through three phases -- a {\em growing phase}, a {\em matured phase} and an {\em interdisciplinary phase}. In the growing phase, the field accumulates ideas from other fields. In the matured phase, the field produces many in-house citations. It also starts receiving citations from other fields. In the interdisciplinary phase, the field receives myriad of citations from other fields. The mutual interactions among many such fields may in turn create a completely new field.

\if{0}

\begin{itemize}


\item The first striking observation is that the papers/fields with highest level of disciplinarity or interdisciplinarity i.e., minimum or maximum values of RDI and KDI respectively,  tend 
to draw less scientific impact than the papers with a balanced mix (i.e., values of RDI and KDI falling in the middle range). This observation is highly contradictory to the traditional belief that more interdisciplinarity perhaps inherently brings a good quality of research. Therefore, we suggest that there should be an optimal level of interdisciplinarity that controls the research not to become diverse and not to remain to focused.

\item The second important observation is the presence of  ``reciprocity'' among the scientific fields based on citation interactions. We
noticed a high correlation between the fraction of incoming and outgoing citations among pair-wise related fields. More
precisely, the fields dealing with similar research areas tend to be confined in producing and consuming citations, thus exhibiting high
correlations. This effect is more prominent among more disciplinarity fields working in same research areas. 

\item  Finally, we studied the evolution of a new research field due to the cross-field knowledge exchange through citation interactions
and observe how nicely the bibliographic indicators are capable of exploring the entire growth dynamics of a field. In a
step further, we observed that in
general a research field passes through three stages in its life -- a growing phase, a matured phase and an interdisciplinary phase.

\end{itemize}
 
 \fi
 
By giving the profiles
of research fields and interdisciplinarity indicators, and other facets
such as interest of the authors, venue of publication, one
can build up a specialized recommendation system aiming to
predict future combination of fields generating new interdisciplinary area of research. It would be interesting to
see how the release of research grants correlates with the evolution of interdisciplinarity. Moreover, as mentioned earlier, the current study on exploring the life trajectory
of a research field needs further investigation with other domains such as physics, biology to obtain a universal signature. We would also be interested to see how the the evolution of a research field can be  modeled theoretically.

\bibliographystyle{spmpsci}

\end{document}